\documentclass[12pt]{article}
\usepackage{epsfig}
\usepackage{graphicx}
\usepackage{amssymb}
\usepackage{amsmath}
\usepackage{amsthm}
\usepackage{amsfonts}
\usepackage{mathrsfs}
\usepackage[dvips]{color}
\usepackage{multirow}

\newcommand{\R}{\mathbb{R}}
\newcommand{\C}{\mathbb{C}}

\newcommand{\fa}{\mathfrak{a}}

\newcommand{\fu}{\mathfrak{u}}

\newcommand{\fn}{\mathfrak{n}}

\newcommand{\fs}{\mathfrak{s}}

\newcommand{\fz}{\mathfrak{z}}

\newcommand{\fK}{\mathfrak{K}}

\newcommand{\bM}{\mathbf{M}}

\newcommand{\bS}{\mathbf{S}}

\newcommand{\cP}{\mathcal{P}}

\newcommand{\cT}{\mathcal{T}}

\newcommand{\be}{\begin{equation}}
\newcommand{\ee}{\end{equation}}
\newcommand{\bea}{\begin{eqnarray}}
\newcommand{\eea}{\end{eqnarray}}
\newcommand{\nn}{\nonumber}

\newcommand{\ed}{\end{document}}

\newcommand{\rx}{{\rm x}}

\newcommand{\rz}{{\rm z}}

\newcommand{\bi}{\begin{itemize}}
\newcommand{\ei}{\end{itemize}}

\newcommand{\bce}{\begin{center}}
\newcommand{\ece}{\end{center}}

\newcommand{\RE}{\,{\rm Re}}
\newcommand{\IM}{\,{\rm Im}}

\begin{document}

\title{Self-dual Spectral Singularities and Coherent Perfect Absorbing Lasers without $\cP\cT$-symmetry}

\author{Ali~Mostafazadeh\thanks{E-mail address:
amostafazadeh@ku.edu.tr, Phone: +90 212 338 1462, Fax: +90 212 338
1559}
\\
Department of Mathematics, Ko\c{c} University,\\
34450 Sar{\i}yer, Istanbul, Turkey}
\date{ }
\maketitle

\begin{abstract}
A $\cP\cT$-symmetric optically active medium that lases at the threshold gain also acts as a complete perfect absorber at the laser wavelength. This is because spectral singularities of $\cP\cT$-symmetric complex potentials are always accompanied by their time-reversal dual. We investigate the significance of $\cP\cT$-symmetry for the appearance of these self-dual spectral singularities. In particular, using a realistic optical system we show that self-dual spectral singularities can emerge also for non-$\cP\cT$-symmetric configurations. This signifies the existence of non-$\cP\cT$-symmetric coherent perfect absorbing (CPA) lasers.
\end{abstract}

\begin{figure}
    \begin{center}
    \includegraphics[scale=.5,clip]{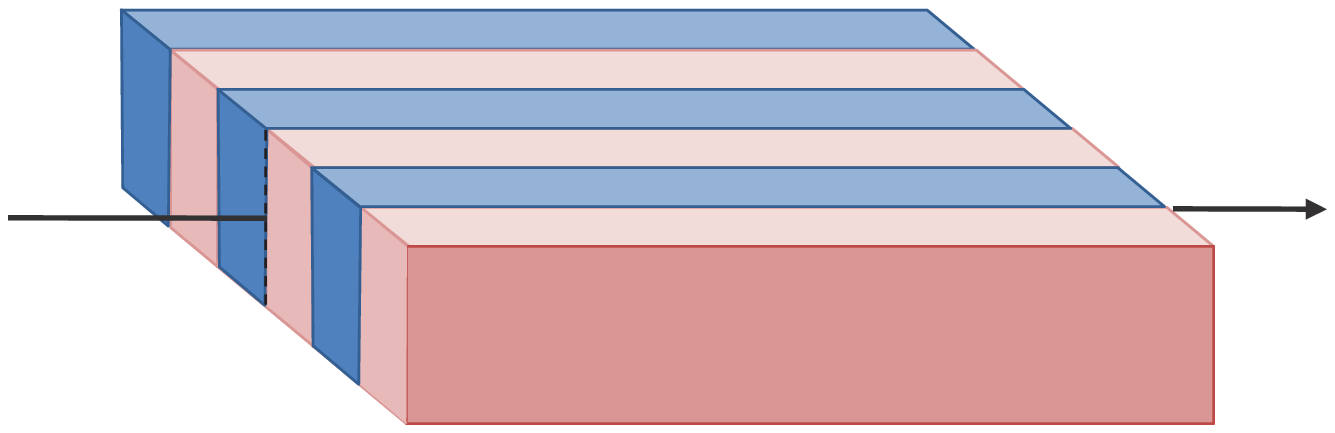}\vspace{-.6cm}
     \caption{(Color online) Schematic view of an active dielectric waveguide with gain/loss properties changing along a normal direction to the longitudinal axis of the guide.\label{fig1}}\vspace{.3cm}
    \includegraphics[scale=.5,clip]{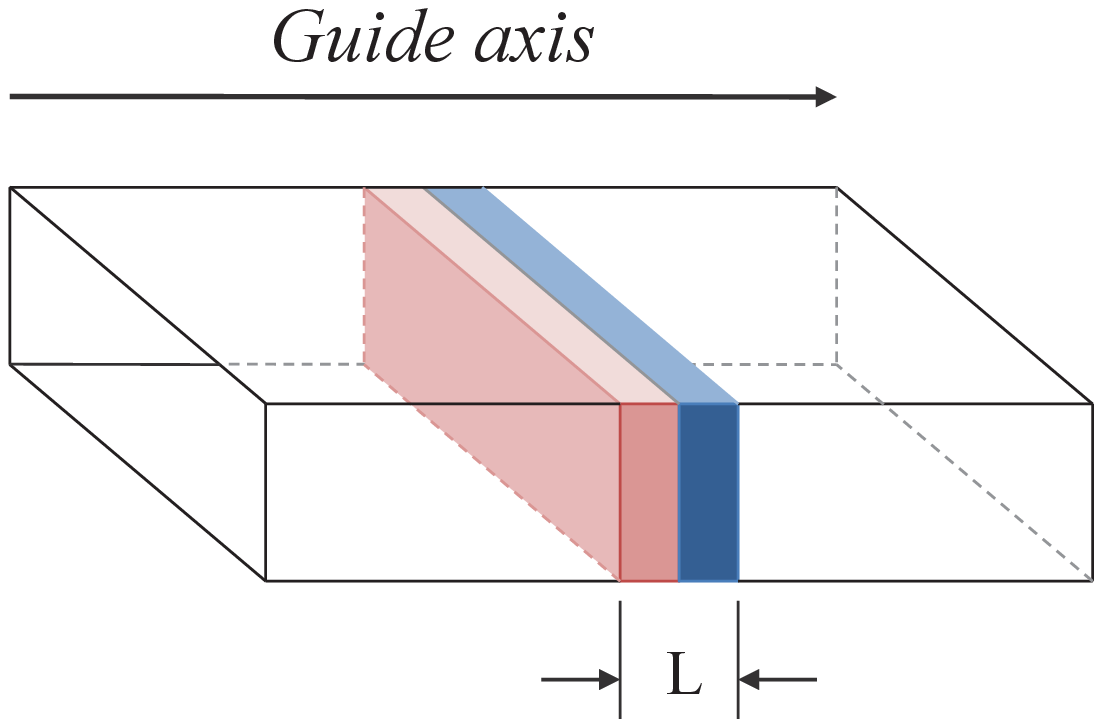}\vspace{-.6cm}
    \caption{(Color online) Schematic view of a waveguide containing an optically active region whose gain/loss properties change along the longitudinal axis of the guide.\label{fig2}}
    \end{center}
    \end{figure}

The recent interest in the role of $\cP\cT$-symmetry in optics was initiated with the theoretical works predicting unusual properties of $\cP\cT$-symmetric optical lattices \cite{waveguide} and the experimental works \cite{waveguide-exp} studying the occurrence and consequences of exceptional points corresponding to the spontaneous breakdown of $\cP\cT$-symmetry \cite{ingrid2}. Exceptional points have been a subject of both theoretical \cite{ep} and experimental \cite{ep-ex} research for more than two decades, but these early studies did not consider systems possessing $\cP\cT$-symmetry. The study of optical realizations of $\cP\cT$-symmetry and its breakdown that is carried out in \cite{waveguide,waveguide-exp} focuses on the behavior of electromagnetic waves propagating in a waveguide consisting of balanced gain and loss regions such that the gain/loss properties of the system varies along a normal direction to the longitudinal axis of the guide as depicted in Figure~\ref{fig1}.
Another recent development is the discovery of an optical realization of the mathematical concept of ``spectral singularity'' \cite{SS} in the form of zero-width resonances \cite{prl-2009}. The physical model used in \cite{prl-2009} to characterize these resonances also involves a $\cP\cT$-symmetric optical waveguide (Figure~\ref{fig2}), but unlike in \cite{waveguide,waveguide-exp} here the gain/loss properties of the optically active region of the guide changes along its longitudinal axis. This particular waveguide configuration was initially proposed in \cite{rdm} as a physical model possessing $\cP\cT$-symmetry. In \cite{jmp-2005} it was used to address the problem of constructing a metric operator \cite{review} for a pseudo-Hermitian Hamiltonian \cite{p123} with a purely continuous spectrum. This in turn paved the way for the discovery of the optical realizations of a spectral singularity \cite{prl-2009}, as the latter appeared as a singularity of the metric operators associated with a delta-function potential with a complex coupling constant \cite{jpa-2006}.

Ref.~\cite{pra-2009} shows that the zero-width optical resonances, that we refer to as optical spectral singularities, can also appear in non-$\cP\cT$-symmetric waveguides. Ref.~\cite{pra-2011a} uses an optical system consisting of a single gain region to show that the mathematical condition for the occurrence of an optical spectral singularity (OSS) coincides with the physical threshold condition for lasing. Therefore, OSSs correspond to lasing at the threshold gain.

At the values of the parameters of the system that a spectral singularity emerges, the transmission coefficient $|T|^2$ as well as both the right and left reflection coefficients, $|R^r|^2$ and $|R^l|^2$, diverge, while the wave number $k$ (and consequently the wavelength $\lambda:=2\pi/k$) remain real \cite{prl-2009}. This is because the (complex) transmission and reflection amplitudes, $T$ and $R^{r,l}$, are related to the entries $M_{ij}$ of the transfer matrix $\bM$ of the system according to
    \be
    T=\frac{1}{M_{22}},~~~~R^l=-\frac{M_{21}}{M_{22}},~~~~~
    R^r=\frac{M_{21}}{M_{22}},
    \label{eq1}
    \ee
and a spectral singularity appears whenever we can satisfy
    \be
    M_{22}=0,
    \label{SS}
    \ee
using a real value of $\lambda$,  \cite{prl-2009}. In this case the system amplifies the background noise and emits radiation of wavelength $\lambda$ from both ends of the scattering region. The time-reversal of this phenomenon corresponds to an optically active region absorbing incident coherent radiation of identical amplitude and phase from both its left- and right-hand sides. This happens at wavelengths where \cite{longhi1}
    \be
    M_{11}=0,
    \label{CPA}
    \ee
and provides the basic principle governing the behavior of a coherent perfect absorber (CPA), also known as an antilaser \cite{antilaser}.

In order to make (\ref{CPA}) more explicit, we invert (\ref{eq1}) and use the fact that the transfer matrix has a unit determinant \cite{jpa-2009} to obtain
    \be
    \bM=\left[\begin{array}{cc}
    T-\displaystyle\frac{R^lR^r}{T} & \displaystyle\frac{R^r}{T}\\ & \\
    \displaystyle-\frac{R^l}{T} & \displaystyle\frac{1}{T}\end{array}\right].
    \label{M}
    \ee
In light of this equation, we can express (\ref{CPA}) in the form
    \be
    T^2=R^l R^r.
    \label{eq4}
    \ee
This corresponds to a zero of (one of the eigenvalues of) the S-matrix of the system,
    \be
    \bS:=\left[\begin{array}{cc}
    T & R^r\\
    R^l & T\end{array}\right],
    \label{S}
    \ee
while (\ref{SS}), that is equivalent to
    \be
    R^l,R^r,T\to\infty,
    \label{eq5}
    \ee
marks a pole of (one of the eigenvalues of) $\bS$, \cite{prl-2009}.

It is easy to see that equations (\ref{eq4}) and (\ref{eq5}) are not contradictory. This shows that there may exist a wavelength $\lambda$ at which one of the eigenvalues of the $S$-matrix diverges while the other vanishes. This corresponds to the situation that a spectral singularity accompanies its time-reversed dual. We refer to such a spectral singularity as a ``self-dual spectral singularity.'' An optical system supporting a self-dual spectral singularity with wavelength $\lambda$ emits electromagnetic radiation at this wavelength unless it is subjected to incident radiation of the same wavelength and identical amplitude and phase from both sides. In the latter case, the system acts as a CPA. Therefore, it is called a ``CPA-laser'' \cite{CGS}.

Self-duality of spectral singularities is a characteristic feature of one-dimensional $\cP\cT$-symmetric scattering potentials. To see this we recall that the transfer matrix of such a potential satisfies, for real values of $k$, $\bM^*=\bM^{-1}$, \cite{longhi1}. Therefore,
    \be
    \RE(M_{12})=\RE(M_{21})=0,~~~~~~~~~~~~~M_{11}^*=M_{22}.
    \label{eq8}
    \ee
In particular, any zero of $M_{22}$ is also a zero of $M_{11}$. This shows that spectral singularities of $\cP\cT$-symmetric potentials are self-dual. Furthermore, one can easily show, using (\ref{M}) and (\ref{eq8}), that for a $\cP\cT$-symmetric scattering potential,
    \bea
    &&e^{2i(\varphi_l-\varphi_t)}=e^{2i(\varphi_r-\varphi_t)}=-1,
    \label{PT-phase}\\
    &&|T|^2\pm|R^lR^r|=1,
    \label{eq3}
    \eea
where $\varphi_l,\varphi_r,\varphi_t$ are respectively the argument of $R^l$, $R^r,T$, i.e., $e^{i\varphi_l}:=R^l/|R^l|$, $e^{i\varphi_r}:=R^r/|R^r|$,  $e^{i\varphi_t}:=T/|T|$, and the unspecified sign in (\ref{eq3}) stands for $e^{i(\varphi_l+\varphi_r-2\varphi_t)}$. Equation~(\ref{eq3}) is the $\cP\cT$-symmetric complex generalization of the standard continuity relation, $|T|^2+|R^{r,l}|^2=1$, that is recently reported in \cite{stone2012}.

\begin{figure}
    \begin{center}
    \includegraphics[scale=.6,clip]{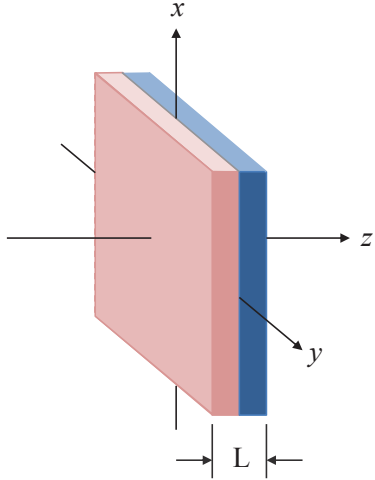}\vspace{-.6cm}
     \caption{(Color online) Schematic view of an infinite planar slab of optically active material that consists of two layers of equal thickness and different loss/gain properties. \label{fig3}}
    \end{center}
    \end{figure}

In this article, we use a simple yet sufficiently rich optical system that allows for a clear assessment of the importance of $\cP\cT$-symmetry for realizing a CPA-laser. Figure~\ref{fig3} shows a schematic description of this system.
It consists of a two-layer infinite planar slab of optically active material. The layers have equal thickness, $L/2$, and different complex refractive indices, $\fn_1$ and $\fn_2$. We examine the linearly polarized time-harmonic electromagnetic waves propagating along the normal direction to the slab. If we use a Cartesian coordinate system whose positive $\rx$- and $\rz$-axes are respectively along the polarization and propagation directions, the wave equation satisfied by the electric field $\vec E(\rz,t)=e^{-i\omega t}\Psi(\rz)\hat e_{\rx}$ reduces to the Helmholtz equation
    \be
    \Psi''(\rz)+k^2\fn(\rz)^2\Psi(\rz)=0,
    \label{H-eq}
    \ee
where $\omega$ and $k:=\omega/c$ are respectively the angular frequency and wave number, $\hat e_{\rx}$ is the unit vector pointing along the positive $\rx$-axis, $\Psi:\R\to\C$ is a continuously differentiable function, and
    \be
    \fn(\rz):=\left\{\begin{array}{ccc}
    \fn_1 & {\rm for} & -\frac{L}{2}\leq \rz <0,\\
    \fn_2 & {\rm for} & 0\leq \rz \leq \frac{L}{2}\\
    1 & {\rm for} & |\rz|>\frac{L}{2}.
    \end{array}\right.
    \label{eq21}
    \ee
Our aim is to examine the consequences of imposing (\ref{SS}) and (\ref{CPA}).

In \cite{pra-2011b,p105} we derive a formula for the transfer matrix of complex potentials vanishing outside a closed interval. In order to make use of this formula, we introduce the dimensionless quantities
    \be
    x:=\frac{\rz}{L}+\frac{1}{2},~~~~~\fK:=Lk,~~~~~
    \psi(x):=\Psi(Lx-\mbox{$\frac{L}{2}$}),
    ~~~~~v(x):=\fK^2\left[1-\fn(Lx-\mbox{$\frac{L}{2}$})^2\right],
    \label{scale}
    \ee
and express (\ref{H-eq}) as the time-independent Schr\"odinger equation:
    \be
    -\psi''(x)+v(x)\psi=\fK^2\psi(x),
    \label{sch-eq}
    \ee
where
    \bea
    v(x)=\left\{\begin{array}{ccc}
    \fK^2(1-\fn_1^2) & {\rm for} & x\in [0,\mbox{$\frac{1}{2}$}),\\
    \fK^2(1-\fn_2^2)  & {\rm for} & x\in [\mbox{$\frac{1}{2}$},1],\\
    0 & {\rm for} & x\notin [0,1].
    \end{array}\right.
    \label{eq22}
    \eea
Note that $v$ is a $\cP\cT$-symmetric potential provided that
$\fn_1^*=\fn_2$.

It is not difficult to show that the transfer matrix of the potential (\ref{eq22}) has the form \cite{p105}:
    \be
    \bM=\frac{1}{2i\fK}\left[\begin{array}{cc}
    -e^{-i\fK}(\Gamma_{1+}-2\Gamma_{2+}) &
    e^{-i\fK} \Gamma_{1+}\\
    e^{i\fK}(\Gamma_{1-}-2\Gamma_{2-}) &
    -e^{i\fK} \Gamma_{1-}
    \end{array}\right],
    \label{Mt=}
    \ee
where
    \be
    \Gamma_{j\pm}:=\phi_j'(1)\pm i\fK\phi_j(1),
    \label{eq30}
    \ee
and $\phi_j$ are the solutions of (\ref{sch-eq}) in the interval
$[0,1]$ that satisfy the initial conditions: $\phi_1(0)=1$, $\phi_1'(0)=-i\fK$,
$\phi_2(0)=1$, and $\phi_2'(0)=0$, \cite{pra-2011b}. They have the following explicit form.
    \be
    \phi_j(x):=\left\{\begin{array}{ccc}
    A_j e^{i\fK\fn_1 x}+B_je^{-i\fK\fn_1 x} & {\rm for} & x\in[0,\mbox{$\frac{1}{2}$}),\\ && \\
    C_je^{i\fK\fn_2 x}+D_je^{-i\fK\fn_2 x} & {\rm for} & x\in[\mbox{$\frac{1}{2}$},1],\end{array}\right.
    \label{phi=}
    \ee
where
    \bea
    A_1&:=&\frac{1}{2}\left(1-\frac{1}{\fn_1}\right),~~~~~
    B_1:=\frac{1}{2}\left(1+\frac{1}{\fn_1}\right),~~~~~
    A_2:=B_2:=\frac{1}{2},
    \label{eq31}\\
    C_1&:=&\frac{e^{-i\fa_2}}{2}\left[\left(1-\frac{1}{\fn_2}\right)\cos\fa_1
    +i\left(\frac{\fn_1}{\fn_2}-\frac{1}{\fn_1}\right)\sin \fa_1\right],
    \label{eq32}\\
    D_1&:=&\frac{e^{i\fa_2}}{2}\left[\left(1+\frac{1}{\fn_2}\right)\cos\fa_1
    -i\left(\frac{\fn_1}{\fn_2}+\frac{1}{\fn_1}\right)\sin \fa_1\right],
    \label{eq33}\\
    C_2&:=&\frac{e^{-i\fa_2}}{2}\left[\cos\fa_1
    +\frac{i \fn_1 \sin \fa_1}{\fn_2}\right],~~~~~
    D_2:=\frac{e^{i\fa_2}}{2}\left[\cos\fa_1
    -\frac{i \fn_1 \sin \fa_1}{\fn_2}\right],
    \label{eq35}
    \eea
and
    \be
    \fa_j:=\frac{\fn_j\fK}{2}.
    \ee

Notice that the change of variable: $\rz\to x=\rz/L+1/2$ involves a shift in the original of the $\rz$ axis that changes the form of the transfer matrix of the system. It is easy to compute its effect and show that the transfer matrix associated with the coordinate system centered at $\rz=0$ is given by
    \be
    \bM=\frac{1}{2i\fK}\left[\begin{array}{cc}
    -e^{-i\fK}(\Gamma_{1+}-2\Gamma_{2+}) &
    \Gamma_{1+}\\
    \Gamma_{1-}-2\Gamma_{2-} &
    -e^{i\fK} \Gamma_{1-}
    \end{array}\right].
    \label{M=}
    \ee
According to this equation, spectral singularities and their time-reversal dual are respectively given by the real values of $\fK$ satisfying
    \bea
    &&\Gamma_{1-}=0,
    \label{SS1}\\
    &&\Gamma_{1+}-2\Gamma_{2+}=0.
    \label{CPA1}
    \eea
In particular the system supports a self-dual spectral singularity and acts as a CPA-laser provided that we can satisfy both these equations for a real $\fK$. Substituting (\ref{phi=}) -- (\ref{eq35}) in (\ref{eq30}), we find after a lengthy set of calculations the following simple expressions for $\Gamma_{1-}$ and $\Gamma_{1+}-2\Gamma_{2+}$.
    \bea
    &&\Gamma_{1-}=\frac{-\fK}{2\fn_1\fn_2}\Big[
    \fn_+\tilde\fn_+\sin\fa_++\fn_-\tilde\fn_-\sin\fa_-+
    i(\fn_+^2\cos\fa_+-\fn_-^2\cos\fa_-)\Big],
    \label{eq41}\\
    &&\Gamma_{1+}-2\Gamma_{2+}=\frac{\fK}{2\fn_1\fn_2}\Big[
    \fn_+\tilde\fn_+\sin\fa_++\fn_-\tilde\fn_-\sin\fa_--
    i(\fn_+^2\cos\fa_+-\fn_-^2\cos\fa_-)\Big],
    \label{eq42}
    \eea
where
    \be
    \fn_\pm:=\fn_1\pm\fn_2,~~~~~\tilde\fn_\pm:=\fn_1\fn_2\pm 1=\frac{1}{4}(\fn_+^2-\fn_-^2)\pm1,~~~~~
    \fa_\pm:=\fa_1\pm\fa_2=\frac{\fn_\pm\fK}{2}.
    \label{eq43}
    \ee

In light of (\ref{eq41}) and (\ref{eq42}), we can easily reduce (\ref{SS1}) and (\ref{CPA1}) to
    \be
    \fn_+\tilde\fn_+\sin\fa_+=-\fn_-\tilde\fn_-\sin\fa_-,~~~~~ \fn_+^2\cos\fa_+=\fn_-^2\cos\fa_-.
    \label{eq51}
    \ee
Squaring both sides of these equations and using the the trigonometric identity,
$\sin^2\fa_\pm+\cos^2\fa_\pm=1$, and the last equation in (\ref{eq43}), we obtain
    \bea
    \cos\left( \frac{\fn_-\fK}{2} \right)&=&
    \pm\frac{i\fn_+\fs}{\fn_-},
    \label{eq52}\\
    \cos\left( \frac{\fn_+\fK}{2} \right)&=&\pm\frac{i\fn_-\fs}{\fn_+},
    \label{eq53}
    \eea
where
    \be
    \fs:=\sqrt{\frac{(\fn_1^2+1)(\fn_2^2+1)}{
    (\fn_1^2-1)(\fn_2^2-1)}}
    =\sqrt{\frac{(\fn_-\tilde\fn_-)^2-(\fn_+\tilde\fn_+)^2}{
    (\fn_-\tilde\fn_+)^2-(\fn_+\tilde\fn_-)^2}}.
    \label{p=}
    \ee
Eqs.~(\ref{eq52}) and (\ref{eq53}) are a pair of complex transcendental equations involving two complex and one real variables, $\fn_\pm$ and $\fK$, respectively.

First, we examine the special case that the refractive indices $\fn_1$ and $\fn_2$ have the same real part, say $n_0$, i.e., they correspond to different amounts of gain/loss achieved using the same active medium. Denoting the imaginary part of $\fn_j$ by $\kappa_j$, so that $\fn_j=n_0+i\kappa_j$, we find $\fn_-=i(\kappa_1-\kappa_2)$ and $\fn_+=2n_0+i(\kappa_1+\kappa_2)$. Inserting these relations in (\ref{eq52}) gives
    \be
    (\kappa_1-\kappa_2)\cosh\left[\frac{(\kappa_1-\kappa_2)\fK}{2}\right]=\pm
    [2n_0+i(\kappa_1+\kappa_2)]\fs.
    \label{cosh=real}
    \ee
Because the left-hand side of this equation is real, so must be its right-hand side. However, notice that for non-exotic material $1\leq n_0 <5$, and $|k_j|$ are at least three orders of magnitude smaller than $n_0$, i.e., $|k_j|\lessapprox 10^{-3}n_0$. Therefore, we can approximate $\fs$ by expanding it in power series in $k_j$ and ignoring the quadratic and higher order terms in the latter. This gives
    \[\fs\approx\left(\frac{n_0^2+1}{n_0^2-1}\right)\left[1+
    \frac{2in_0(\kappa_1+\kappa_2)}{n_0^4-1}\right].\]
Now, we insert this relation in (\ref{cosh=real}) and demand that its right-hand side be real. The result is
    \be
    (\kappa_1+\kappa_2)\left(1+\frac{4n_0^2}{n_0^4-1}\right)\approx 0.
    \label{cosh=real2}
    \ee
Because $n_0\geq 1$,
    \[1+\frac{4n_0^2}{n_0^4-1}\geq 1+\frac{4}{n_0^2}>1.\]
Therefore, (\ref{cosh=real2}) implies that $\kappa_1+\kappa_2\approx 0$. This means that within the limits set by the physical considerations, the condition that the system supports a self-dual optical spectral singularity implies that $\fn_1\approx\fn_2^*$, i.e., the system possesses $\cP\cT$-symmetry. However, there are more general situations where $\fn_1$ and $\fn_2$ have different real parts and we can still satisfy (\ref{eq52}) and (\ref{eq53}). These correspond to non-$\cP\cT$-symmetric self-dual spectral singularities that we wish to characterize.

We can solve Eqs.~(\ref{eq52}) and (\ref{eq53}) for $\fK$ and use the identity, $\cos^{-1}(\fz)=\pm i\ln(\fz\pm\sqrt{\fz^2-1})$, to express the result in the form:
    \bea
    \fK&=&\pm\frac{2i}{\fn_+} \Big\{\ln\left|\fu_-\right|+
    i\left[{\rm arg}(\pm i\,\fu_-)-2\pi m_-\right]\Big\},
    \label{eq71}\\
    \fK&=&\pm\frac{2i}{\fn_-} \Big\{\ln\left|\fu_+\right|+
    i\left[{\rm arg}(\pm i\,\fu_+)-2\pi m_+\right]\Big\}.
    \label{eq72}
    \eea
Here we have introduced
    \be
    \fu_-:=\sqrt{\frac{\fn_-^2\fs^2}{\fn_+^2}+1}\pm\frac{\fn_-\fs}{\fn_+},~~~~~
    \fu_+:=\sqrt{\frac{\fn_+^2\fs^2}{\fn_-^2}+1}\pm\frac{\fn_+\fs}{\fn_-},
    \label{u-pm}
    \ee
and employed the identity $\ln\fz=\ln|\fz|+i[{\rm arg}(\fz)-2\pi m]$, ${\rm arg}(\fz)$ stands for the principal argument of the complex number $\fz$ (that takes values between $-\pi$ and $\pi$), and $m$ and $m_\pm$ are arbitrary integers.

Notice that although $\fK$ is a real variable, (\ref{eq71}) and (\ref{eq72}) are actually complex equations. In particular, $\fK$ is equal to the real part of the right-hand side of these equations, and their imaginary part must vanish. Therefore, if we denote by $\eta_j$ and $\kappa_j$ the
real and imaginary part of $\fn_j$, for $j=1,2$, so that $\fn_j=\eta_j\pm
i\kappa_j$, and let
    \bea
    \eta_\pm&:=&\RE(\fn_\pm)=\RE(\fn_1)\pm\RE(\fn_2)=\eta_1\pm\eta_2,
    \label{eta-pm}\\
    \kappa_\pm&:=&\IM(\fn_\pm)=\IM(\fn_1)\pm\IM(\fn_2)=\kappa_1\pm\kappa_2,
    \label{kappa-pm}\\
    \varphi_-&:=&{\rm arg}(\pm i\,\fu_-),~~~~~
    \varphi_+:={\rm arg}(\pm i\,\fu_+),
    \label{phi-pm}
    \eea
we find
    \bea
    &&\fK=\pm 2 |\fn_+|^{-2} \big[\kappa_+\ln|\fu_-|+
    \eta_+(2\pi m_--\varphi_-)\big],
    \label{eq73}\\
    &&\fK=\pm 2 |\fn_-|^{-2} \big[\kappa_-\ln|\fu_+|+
    \eta_-(2\pi m_+-\varphi_+)\big],
    \label{eq74}\\
    &&\eta_+\ln|\fu_-|-\kappa_+(2\pi m_--\varphi_-)=0,
    \label{eq75}\\
    &&\eta_-\ln|\fu_+|-\kappa_-(2\pi m_+-\varphi_+)=0.
    \label{eq76}
    \eea
We can solve for $\ln|\fu_\pm|$ in the last two equations and substitute the result in the first two to obtain:
    \be
    \fK=\pm\frac{2\left(2\pi m_--\varphi_-\right)}{\eta_+},
    ~~~~~~~
    \fK=\pm\frac{2\left(2\pi m_+-\varphi_+\right)}{\eta_-}.
    \label{eq77}
    \ee
Because for realistic situations, $1<\eta_+<10$, $0<|\eta_-|<10$,
$|\varphi_\pm|\leq\pi$ and $\fK>10^2$, (\ref{eq77}) implies that
$\pm m_-\gg 1$ and $\pm{\rm sgn} (\eta_-)m_+\gg 1$, where ${\rm sgn}(\eta_-)$ stands for the sign of $\eta_-$. In view of these relations,
we can redefine $m_\pm$ so that $m_\pm >0$ and express (\ref{eq77}) as
    \bea
    \fK=\frac{2\left(2\pi m_-\mp\varphi_-\right)}{\eta_+},
    \label{eq78}\\
    \fK=\frac{2\left(2\pi m_+\mp\varphi_+\right)}{|\eta_-|}.
    \label{eq79}
    \eea
In terms of the new $m_\pm$, (\ref{eq75}) and (\ref{eq76}) take the form
    \bea
    \kappa_-&=&\pm\frac{|\eta_-|\ln|\fu_+|}{2\pi m_+\mp\varphi_+},
    \label{eq80}\\
    \kappa_+&=&\pm\frac{\eta_+\ln|\fu_-|}{2\pi m_-\mp\varphi_-}.
    \label{eq81}
    \eea

Because $m_\pm\gg 1\geq |\varphi_\pm/(2\pi)|$, we can neglect terms proportional to $\varphi_\pm/(2\pi m_\pm)$ in (\ref{eq78}) -- (\ref{eq81}). This gives
    \bea
    \fK&\approx&\frac{4\pi m_-}{\eta_+},
    \label{eq78p}\\
    \fK&\approx&\frac{4\pi m_+}{|\eta_-|},
    \label{eq79p}\\
    \kappa_-&\approx&\pm\frac{|\eta_-|\ln|\fu_+|}{2\pi m_+},
    \label{eq80p}\\
    \kappa_+&\approx&\pm\frac{\eta_+\ln|\fu_-|}{2\pi m_-}.
    \label{eq81p}
    \eea
Equations (\ref{eq78p}) and (\ref{eq79p}) imply the curious relation:
    \be
    \frac{|\eta_-|}{\eta_+}\approx\frac{m_+}{m_-}.
    \label{eq82}
    \ee
In particular, because $|\eta_-|<\eta_+$, we have $m_+<m_-$.

Next, we note that for realistic models $|\kappa_\pm| \lessapprox 10^{-3}$ and $\eta_+>1$. Therefore, if we consider the cases where $|\eta_-|>0.1$, we have $|\kappa_\pm|\ll|\eta_\pm|$. Neglecting terms of order $|\kappa_\pm/\eta_\pm|$ in (\ref{u-pm}), we then find
    \bea
    &&\fs\approx\sigma:=\sqrt{\frac{(\eta_1^2+1)(\eta_2^2+1)}{
    (\eta_1^2-1)(\eta_2^2-1)}}=
    \sqrt{\frac{[(\eta_++\eta_-)^2+4][(\eta_+-\eta_-)^2+4]}{
    [(\eta_++\eta_-)^2-4][(\eta_+-\eta_-)^2-4]}},
    \nn\\
    &&|\fu_-|\approx  \sqrt{\frac{\sigma^2\eta_-^2}{\eta_+^2}+1}\pm\frac{\sigma\eta_-}{\eta_+},~~~~~~~~~~
    |\fu_+|\approx  \sqrt{\frac{\sigma^2\eta_+^2}{\eta_-^2}+1}\pm\frac{\sigma\eta_+}{\eta_-}.\nn
    \eea
These relations together with (\ref{eq78p}) and (\ref{eq79p}) allow us to write (\ref{eq80p}) and (\ref{eq81p}) in the form
    \bea
    \kappa_-&\approx&\pm \frac{|\eta_-|}{2\pi m_+}
    \ln\left(\sqrt{\frac{\sigma^2\eta_+^2}{\eta_-^2}+1}\pm\frac{\sigma\eta_+}{\eta_-}\right)
    \approx \pm 2\, \fK^{-1}
    \ln\left(\sqrt{\frac{\sigma^2\eta_+^2}{\eta_-^2}+1}\pm\frac{\sigma\eta_+}{\eta_-}\right),
    \label{eq91}\\
    \kappa_+&\approx&\pm\frac{\eta_+}{2\pi m_-}
    \ln\left(\sqrt{\frac{\sigma^2\eta_-^2}{\eta_+^2}+1}\pm\frac{\sigma\eta_-}{\eta_+}\right)
    \approx \pm 2\, \fK^{-1} \ln\left(\sqrt{\frac{\sigma^2\eta_-^2}{\eta_+^2}+1}\pm\frac{\sigma\eta_-}{\eta_+}\right),
    \label{eq92}
    \eea
where the pair of unspecified signs appearing in the right-hand side of these equations are unrelated.

The approximate equations (\ref{eq78p}), (\ref{eq79p}), (\ref{eq91}) and (\ref{eq92}) can be used to express $m_\pm$ and $\kappa_\pm$  in terms of $\eta_1$, $\eta_2$ and $\fK$ up to sixteen sign ambiguities that we eliminate by enforcing (\ref{eq51}) and using physical considerations. The resulting approximate values can be the basis for more accurate numerical solutions of (\ref{eq51}). Table~\ref{table1} gives the result of three sample calculations. The property that these configurations come in complex-conjugate pairs is a manifestation of the fact that the time-reversal transformation maps each solution of the equations determining self-dual spectral singularities, i.e., $M_{11}=M_{22}=0$, to a solution of these equations with the same value of the wave number $k$.\footnote{This is true for every scattering potential.} As we see from Table~\ref{table1}, our simple approximation scheme produces values which are in a very good agreement with the exact (numerical) results.
     \begin{table}
     \begin{center}
        \begin{tabular}{|c|c|c|} \hline
        & Approximate Values\ & Exact (Numerical) Values \\ \hline \hline
        $\fn_1$ & $3.600\pm1.180\times10^{-3}i$ & $3.603\pm1.178\times
        10^{-3}i$\\
        \hline
        $\fn_2$ & $1.500\mp2.241\times10^{-3}i$ & $1.498\mp2.243\times10^{-3}i$
        \\
        \hline \hline
        $\fn_1$ & $3.600\pm2.524\times10^{-3}i$ & $3.600\pm2.520\times10^{-3}i$\\
        \hline
        $\fn_2$ & $3.000\mp2.698\times10^{-3}i$ & $2.997\mp2.695\times10^{-3}i$\\
         \hline \hline
         $\fn_1$ & $3.000\pm1.372\times10^{-3}i$ & $3.000\pm1.370\times10^{-3}i$\\
        \hline
        $\fn_2$ & $1.400\mp2.429\times10^{-3}i$ & $1.398\mp2.431\times10^{-3}i$\\
        \hline
        \end{tabular}
        \caption{Results of some typical values for the parameters of the system that yield a non-$\cP\cT$-symmetric CPA-laser operating for $\fK=400\pi$. This corresponds to $L/\lambda=200$. The exact values have been obtained by a numerical treatment of (\ref{eq51}) after fixing the value of $\fK$.
        \label{table1}}
        \end{center}
     \end{table}

In conclusion, in this article we have performed an analytic treatment of the problem of finding self-dual spectral singularities of an experimentally accessible model. Self-duality is a characteristic feature of the spectral singularities of complex $\cP\cT$-symmetric scattering potentials. However, as our explicit calculations show, one can also achieve a self-dual spectral singularity for a non-$\cP\cT$-symmetric potential. In terms of the optical realizations of spectral singularities, this signifies the existence of a large class of non-$\cP\cT$-symmetric CPA-lasers. These may be of practical interest, for these devices are not bound by the strict requirement of having balanced gain and loss components.\footnote{A scattering potential involving $n$ independent complex coupling constants has $n-1$ degrees of freedom, if we demand that it is $\cP\cT$-symmetric and has a self-dual spectral singularity. This number is $2n-3$, if we relax the requirement of $\cP\cT$-symmetry. For a proof, see the appendix.}

\vspace{3mm}
\noindent \textbf{{Acknowledgments:}} This work has been supported by  the Scientific and Technological Research Council of Turkey (T\"UB\.{I}TAK) in the framework of the project no: 110T611, and by the Turkish Academy of Sciences (T\"UBA). I wish to thank Aref Mostafazadeh for his help in preparing the figures.

\section*{Appendix}

In order to give an idea of the amount of freedom we gain by relaxing the condition of $\cP\cT$-symmetry in devising a CPA-laser, consider a system that can be modeled using a complex scattering potential involving $n$ independent complex coupling constants, $\fz_1,\fz_2,\cdots,\fz_n$. We can write it in the form: $v_n(x)=\sum_{\ell=1}^m\fz_\ell f^+_\ell(x)+i\sum_{\ell=m+1}^n\fz_\ell f^-_\ell(x)$, where $f^+_\ell:\R\to\R$ and $f^-_\ell:\R\to\R$ are respectively even and odd real-valued functions that form a linearly-independent subset of the vector space of all real-valued functions defined on $\R$.

Requiring that $v_n$ be $\cP\cT$-symmetric implies that all the coupling constants are real. In this case, $M_{11}=M_{22}^*$, and the condition that $v_n$ has a spectral singularity, which is necessarily self-dual, means to choose these $n$ real coupling constants and the wave number $k$ such that $M_{22}=0$. Because this is a complex equation, it restricts two of the these $n+1$ real parameters. Therefore, imposing $\cP\cT$-symmetry and the condition of the existence of a self-dual spectral singularity amounts to fixing the $n+2$ of the initial $2n+1$ real parameters of the problem, and leaves us with $n-1$ degrees of freedom.

Now, we relax the condition that $v_n$ is $\cP\cT$-symmetric and only demand that it has a self-dual spectral singularity. This means that the coupling constants $\fz_\ell$ and the wave number $k$ satisfy $M_{11}=0$ and $M_{22}=0$. These are a pair of generically independent complex equations. They restrict $4$ of the initial $2n+1$ real physical parameters: $\RE(\fz_\ell)$, $\IM(\fz_\ell)$ and $k$. Therefore, in this case we are left with $2n-3$ degrees of freedom. This is larger than that of the $\cP\cT$-symmetric case, if $n>2$. For the simple model that we consider in this paper, $n=2$ and the number of degrees of freedom left after imposing the condition of the existence of a self-dual spectral singularity is the same for both the $\cP\cT$-symmetric and non-$\cP\cT$-symmetric configurations. However, even in this case the range of values of the physical parameters supporting a self-dual spectral singularity can be altered almost at will for non-$\cP\cT$-symmetric configurations. Specifically, we can freely choose $\eta_1$, $\eta_2$, and $\fK$ in our approximation scheme and find exact values of $\eta_1$ and $\eta_2$ in a close vicinity of our initial choice for these quantities. For the $\cP\cT$-symmetric case, we can only choose $\eta=\eta_1$ and $\fK$. Therefore, although relaxing the requirement of $\cP\cT$-symmetry does not increase the degrees of freedom of the system, it provides an extra freedom in the choice of the range of values of one of the physical parameters. For $n>2$, non-$\cP\cT$-symmetric configurations (optical systems) supporting a spectral singularity (functioning as a CPA-laser) have a larger number of free parameters. Therefore, they should be practically more favorable over their $\cP\cT$-symmetric counterparts.

\ed
\begin{thebibliography}{99}

\bibitem{waveguide}
Z.~H.~Musslimani, K.~G.~Makris, R.~El-Ganainy,
and D.~N.~Christodoulides, Phys.\ Rev.\ Lett.~{\bf 100}, 030402
(2008); K.~G.~Makris, R.~El-Ganainy, D.~N.~Christodoulidesand and
Z.~H.~Musslimani, Phys.\ Rev.\ Lett.~{\bf 100}, 103904 (2008);
S.~Longhi, Phys.\ Rev.\ Lett.~{\bf 103}, 123601 (2009).

\bibitem{waveguide-exp}
A.~Guo, G.~J.~Salamo, D.~Duchesne, R.~Morandotti, M.~Volatier-Ravat and V.~Aimez,   Phys.\ Rev.\ Lett.~{\bf 103}, 093902 (2009); C.~E.~Rueter, K.~G.~Makris, R.~El-Ganainy, D.~N.~Christodoulides, M.~Segev, D.~Kip, Nature Phys.\ {\bf 6}, 192 (2010).

\bibitem{ingrid2} I.~Rotter, J.~Opt.\ {\bf 12}, 065701 (2010).

\bibitem{ep} W.~D.~Heiss and A.~L.~Sannino, J.~Phys.\ A~{\bf 23},
1167 (1990); W.~D.~Heiss, Phys.\ Rep.~{\bf 242}, 443 (1994);
W.~D.~Heiss, M.~M\"uller, I.~Rotter, Phys.\ Rev.~E {\bf 58},
2894 (1998); E.~Narevicius and N.~ Moiseyev, Phys.\ Rev.\ Lett.~{\bf
81}, 2221 (1998) and {\bf 84}, 1681 (2000); W.~D.~Heiss, Phys.\ Rev.~E {\bf 61}, 929 (2000); I.~Rotter, Phys.\ Rev.~E {\bf 67}, 026204 (2003);
W.~D.~ Heiss,  J.\ Phys.\ A., {\bf 37}, 2455 (2004); A.~P.~Seyranian, O.~N.~Kirillov, and A.~A.~Mailybaev, J.~Phys.~A {\bf 38}, 1723 (2005);
A.~A.~Mailybaev, O.~N.~Kirillov, and A.~P.~Seyranian, Phys.\ Rev.~A {\bf 72}, 014104 (2005); M.~M\"uller and I.~Rotter, J.~Phys.~A {\bf 41}, 244018 (2008); H. Mehri-Dehnavi and A. Mostafazadeh, J.~Math.\ Phys.\ {\bf 49}, 082105 (2008).

\bibitem{ep-ex} C. Dembowski, H.-D. Gr\"{a}f, H. L. Harney, A. Heine, W. D. Heiss, H. Rehfeld, and A. Richter, Phys.\ Rev.\ Lett., {\bf 86}, 787 (2001); C.~Dembowski, B. Dietz,  H.-D. Gr\"{a}f, H. L. Harney, A. Heine, W. D. Heiss, and A. Richter, Phys.\ Rev.\ E., {\bf 69}, 056216 (2004); T.~Stehmann, W.~D.~Heiss, and F.~G.~Scholtz, J.~Phys.~A {\bf 37}, 7813 (2004).

\bibitem{SS} M.~A.~Naimark, Trudy Moscov.\ Mat.\ Obsc.\ \textbf{3}, 181 (1954) in Russian, English translation: Amer.\ Math.\ Soc.\ Transl.\
(2), \textbf{16}, 103 (1960); R.~R.~D.~Kemp, Canadian J. Math.
\textbf{10}, 447 (1958); J.~Schwartz, Comm.\ Pure Appl.\ Math.
\textbf{13}, 609 (1960); G.~Sh.~Guseinov, Pramana.\ J.~Phys.\
\textbf{73}, 587 (2009).

\bibitem{prl-2009} A.~Mostafazadeh, Phys.\ Rev.\ Lett.~\textbf{102}, 220402
(2009).

\bibitem{rdm} A. Ruschhaupt, F.~Delgado, and J.~G.~Muga, J.~Phys.~A
{\bf 38}, L171 (2005).

\bibitem{jmp-2005} A.~Mostafazadeh, J.~Math.~Phys.~{\bf 46}, 102108 (2005) and {\bf 47}, 072103 (2006).

\bibitem{review} A.~Mostafazadeh, Int.\ J.~Geom.\ Meth.\ Mod.\
Phys.~\textbf{7}, 1191 (2010); arXiv:0810.5643.

\bibitem{p123} A.~Mostafazadeh, J.\ Math.\ Phys.\ {\bf 43}, 205, 2814, and
3944 (2002).

\bibitem{jpa-2006} A.~Mostafazadeh, J.\ Phys.~\textbf{39}, 13495 (2006).

\bibitem{pra-2009} A.~Mostafazadeh, Phys.\ Rev.\ A \textbf{80}, 032711
(2009).

\bibitem{pra-2011a} A.~Mostafazadeh, Phys. Rev.~A \textbf{83}, 045801 (2011).

\bibitem{longhi1} S.~Longhi,  Phys.\ Rev.\ A  \textbf{82}, 031801 (2010).

\bibitem{antilaser} Y.~D.~Chong, L.~Ge, H.~Cao, and A.~D.~Stone, Phys.\ Rev.\ Lett.\ {\bf 105}, 053901 (2010);
    W.~Wan, Y.~Chong, L.~Ge, H.~Noh, A.~D.~Stone, and H.~Cao, Science 331, 889 (2011);
   S.~Longhi,  Phys.\ Rev.\ A  \textbf{82}, 031801 (2010),  Phys.\ Rev.\ A  \textbf{83}, 055804 (2011), and Phys.\ Rev.\ Lett.~\textbf{107}, 033901 (2011);
    L.~Ge, Y.~D.~Chong,, S.~Rotter, H.~E.~T\"ureci, and A.~D.~Stone, Phys.\ Rev.\ A~\textbf{84}, 023820 (2011).

\bibitem{CGS} Y.~D.~Chong, L.~Ge, and A.~D.~Stone, Phys.\ Rev.\ Lett.~\textbf{106}, 093902 (2011).

\bibitem{jpa-2009} A.\ Mostafazadeh and H.\ Mehri-Dehnavi , J. Phys. A \textbf{42}, 125303 (2009).

\bibitem{stone2012} L.~Ge, Y.~D.~Chong, and A.~D.~Stone, Phys.\ Rev.~A {\bf 85}, 023802 (2012).

\bibitem{pra-2011b} A.~Mostafazadeh, Phys.\ Rev.\ A \textbf{84}, 023809
(2011).

\bibitem{p105} A.~Mostafazadeh and S.~Rostamzadeh, preprint arXiv:1204.2701.


\end{thebibliography}
